\documentclass[12pt,aps,superscriptaddress,twocolumn,reprint]{revtex4-1} 
\usepackage{amssymb,amsmath}
\usepackage{graphicx}
\begin{document}

% Title of the article
\title{Competing Structures %and Lattice Dynamics 
       in (In,Ga)Se and (In,Ga)$_2$Se$_3$ Semiconductors}

% Authors
\author{J.~Srour}
%\affiliation{Universit\'e de Lorraine, LCP-A2MC, 1 Bd Arago, F-57078 Metz, France}
%\affiliation{Universit\'e Libanaise, Facult\'e des Sciences, Beirut, Lebanon}

\author{A.~Postnikov}
\email{andrei.postnikov@univ-lorraine.fr}
\author{M.~Badawi}
\affiliation{Universit\'e de Lorraine, LCP-A2MC, 1 Bd Arago, F-57078 Metz, France}

\author{F.~El Haj Hassan}
\affiliation{Universit\'e Libanaise, Facult\'e des Sciences, Beirut, Lebanon}

%E-mail-address of corresponding author
%\mail{e-mail
%  \textsf{andrei.postnikov@univ-lorraine.fr}, Phone:
%  +33-387-315873, Fax: +33-387-315801}

% author's affiliations/addresses

% Please select about four verbal keywords for your manuscript.
\keywords{Semiconductors, layer structures, ab initio, electronic structure}

\begin{abstract}
The electronic structure of four polytypes 
($\beta$, $\gamma$, $\delta$ and $\varepsilon$)
of hexagonal GaSe and InSe is calculated
from first principles, using the WIEN2k and VASP codes and PBEsol prescription
for the exchange-correlation potential, aiming specifically at elucidating the crystallographic
parameters and comparing the energy placement of corresponding competing structures.

Further on, the compounds with different composition of the same constituents,
namely the ordered-vacancies systems Ga$_2$Se$_3$ and In$_2$Se$_3$, were subject to
a similar study, of which the relaxed crystal structure data for three different
phases is reported. Comparison is done with the nominal wurtzite structure
over which the cation vacancies are introduced, and the relaxation pattern discussed.
\end{abstract}

\maketitle   % please do not remove

\section{Introduction}
The III-VI semiconductors, InSe and GaSe, exist as stoichiometric binaries
in peculiar structures whose common element is the double layer,
(VI)-(III)-(III)-(VI), that includes a cation-cation bond perpendicular to the layer.
Otherwise, cations are in tetrahedral environment (comprising three anions 
at the double layer's surface), and anions share three covalent bonds
with neighboring cations. Different assembling of double layers gives rise
to a number of polytypes; e.g., Kuhn \emph{et al.} \cite{PSSA31-469}
identified $\beta$, $\varepsilon$, $\gamma$ and $\delta$ for GaSe.
For InSe, the rhombohedral $\gamma$ phase seems to be predominant,
according to Likforman and Guittard \cite{ComRenAcadSciC279-33}.
Some of the polytypes have been studied in first-principles calculations,
according to experimental availability and need for interpretation --
see, e.g., Adler \emph{et al.} \cite{PRB57-3726}
and Zhang \emph{et al.} \cite{ChinPhysLett23-1876} for GaSe
and Gomes da Costa \emph{et al.} \cite{PRB48-14135} for InSe.
As is common with hexagonal polytypes, their relative stabilities 
are very close; however, it is noteworthy that the most prominent phases are
markedly different for GaSe and InSe.
We perform a systematic study of electronic properties and relaxed structures,
some parts of which have been previously reported \cite{PSSB253-1472}.
In the present work, we concentrate on the total energy results
concerning the phase diagram
of polytypes.

\section{Calculation setup}
First-principles calculations of electronic structure and equilibrium
crystallographic parameters were done using two methods in comparison,
the all-electrons WIEN2k package \cite{wien2k} and the plane-wave 
VASP package \cite{vasp} %,PRB47-558}
which employs the plane-wave basis set in combination with the projected augmented
waves (PAW) technique to treat the core states. With both methods, a number of
exchange-correlation (XC) schemes was used; the present contribution 
addresses the results obtained with the PBEsol scheme \cite{PRL100-136406},
which allows to compare results of two calculation codes.

\begin{figure}[t]%
\includegraphics*[width=\linewidth]{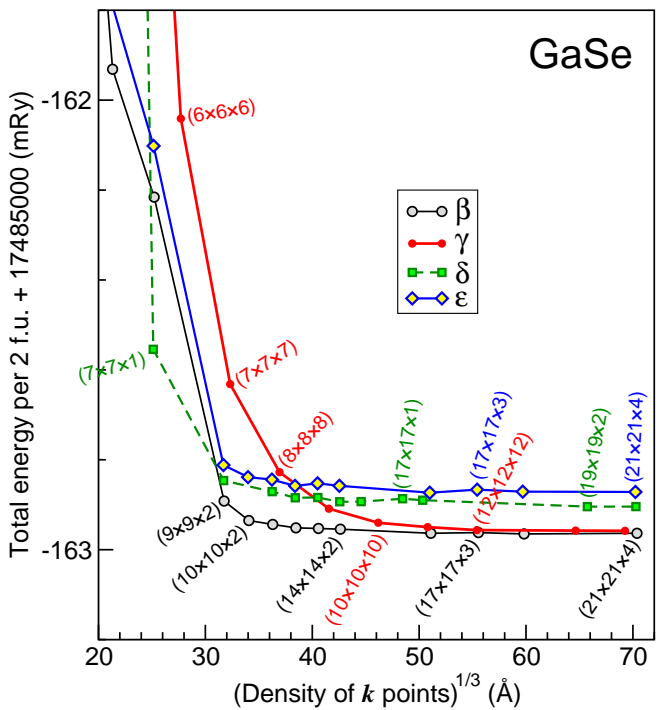}
\caption{\label{fig:k-mesh}
Total energy varying with the density of $\mathbf{k}$ points,
for different polytypes of GaSe, after the WIEN2k calculations.
See text for details.}
\end{figure}

The VASP code offers the PAW database for different elements; the WIEN2k method
is effectively ``exact'' (at the level of DFT in combination with the given XC 
approximation), since the (small) number of technical parameters, which may affect
the accuracy of calculations, can be systematically tested. In the present study,
our ambition was to reliably resolve the relative stability of different polytypes
of GaSe and InSe, that is, to analyse their energy-volume curves. Due to a tiny difference
between the energies of polytypes (of the order of 1 meV per two formula units, i.e., 4 atoms),
we had to press for utmost precision, increasing and testing the two sensitive cutoffs 
present, in a comparable way, in both calculation methods. These are the planewave cutoff
(for the size of the basis set and for the Fourier transform of the density) 
and the density of $\mathbf{k}$-points used for integration over the Brillouin zone (BZ).
Fig.~\ref{fig:k-mesh} shows the calculated (with the WIEN2k code)
total energy for several polytypes
of GaSe, at some fixed structure (near the equilibrium one) for each polytype,
varying in dependence of the number of divisions along the edges of the reciprocal
unit cell (RUC), used for tetrahedron integration. The total energy consistently lowers
as the integration becomes more accurate, but this occurs not at the same rate
for different polytypes. The average linear density of $\mathbf{k}$-points
was taken as a uniform parameter controlling  
the accuracy of the $\mathbf{k}$-space integration,
whatever the shape of the BZ. One notes that
the relative energy placement of hexagonal phases ($\beta$, $\delta$, $\varepsilon$)
gets stabilized already at 
$\mathbf{k}$-mesh with 8 divisions along the planar dimensions of the (flat) RUC.
However, the rhombohedral $\gamma$ phase possesses the RUC                                                                                                                                                                                                                                                                                                                                                                                                                                                                                                                                                                                                                                                                                                                                                                                                                                                                                                                                                                                                                                                                                                                                                                                                                                                                                                                                                                                                                                                                                                                                                                                                                                                                                                                                                                                                                                                                                                                                                                                                                                                                                                                                                                                                                                                                                                                                                                                                                                                                                                                                                                                                                                                                                                                                                                                                                                                                                                                                                                                                                                                                                                                                                                                                                                                                               
of different shape, split into tetrahedra in different way. Consequently,
the cancellation of the systematic integration error does not happen so readily.
One sees from Fig.~\ref{fig:k-mesh} that the total energy of the $\gamma$ phase stabilizes, 
relative to the energies of other phases and to the accuracy needed to reliably
distinguish them,
as the number of divisions along the RUC edges exceeds
$12{\times}12{\times}12$. For hexagonal structures, the comparable density of
$\mathbf{k}$ points is achieved at about $20{\times}20{\times}3$
$\mathbf{k}$-divisions along the hexagonal reciprocal vectors.

While performing VASP calculations, we had to be careful to specify (much) stricter
values of thresholds responsible for the accuracy of total energy convergence
and for the stability of calculated forces than it is usually the case. Running 
the calculations with ``nearly default settings'' of VASP lead to very ``noisy''
energy/volume curves which did not allow to reliably discriminate the polytypes
by their energy. 

In practical sense, the structure optimization in VASP is more flexible than in WIEN2k
and allows many options to combine constraining certain degrees of freedom
with relieving the others. The WIEN2k lattice relaxation scenario, which technically
decouples $(i)$ optimization of internal coordinates (following forces on atoms)
for fixed lattice parameters from $(ii)$ exploring the total energy depending on these
parameters, demands a considerable amount of work to be done almost by hand,
via several iterations, in order to arrive at unconstrained optimized structure.

\begin{figure*}[t]%
\includegraphics*[width=0.9\linewidth]{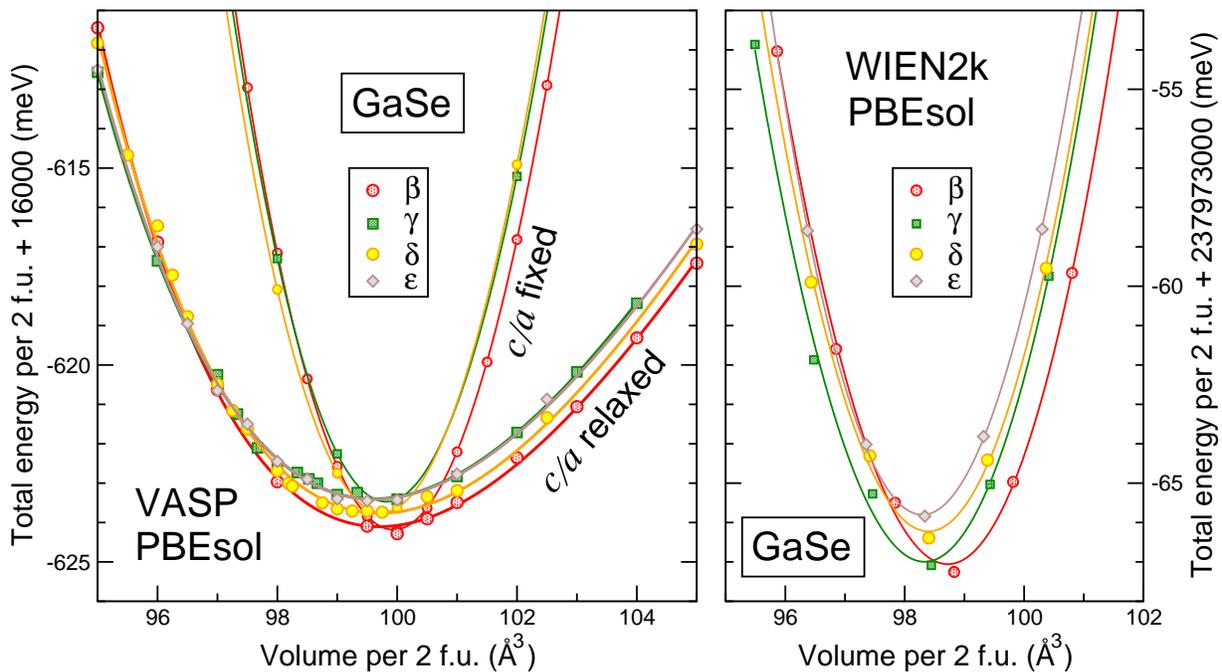}
\caption{\label{fig:GaSe_EneVol}
Energy/volume curves as calculated by VASP (left panel) and WIEN2k (right panel)
for different phases of GaSe, using the PBEsol XC potential.
The steeper set of curves (those for WIEN2k,
and those labelled ``$c/a$ fixed'' for VASP) correspond to the uniform
scaling of the structure at equilibrum.
The softer set of curves (labelled
``$c/a$ relaxed'' for VASP) correspond to the fully relaxed structure for every volume.}
\end{figure*}

\section{Structure optimization of G\lowercase{a}S\lowercase{e}
 and I\lowercase{n}S\lowercase{e}}
In order to provide a meaningful comparison between the results by VASP and WIEN2k,
we show in Fig.~\ref{fig:GaSe_EneVol} the scans of total energy as function of unit cell volume,
which pass through the fully optimized structure for each respective polytype
and explore the \emph{uniform} expansion / compression of the unit cell 
from this optimized geometry. The absolute total energy values do not have any special
meaning, but the relative energy scale makes sense and should be comparable between
the methods. Indeed, the general shape of the nearly parabolic scan in the right panel
depicting the WIEN2k results quite perfectly coincides with the steep parabola
in the left panel, marked ``$c$/$a$ fixed'', which reveals the same type of distortion
applied in VASP calculation. Unfortunately, the agreement between the methods
in what regards fine details, i.e., the placement of individual parabolas, is not that
perfect. We can systematise the less trivial observations as follows:
%\begin{itemize}
%\item
\\ \indent $-$
According to both calculations, the $\beta$ phase is the ground-state one,
producing the lowest-lying parabola.
%\item
\\ \indent $-$
The spread in energy between the polytypes, for a given volume, is very narrow --
around 1~meV (per two formula units, f.u.) in VASP and around 1.5~meV accoring to WIEN2k.
%\item
\\ \indent $-$
The equilibrium volume, not much different throughout the polytypes, is $\sim$100~{\AA}$^3$
(per two f.u.) according to VASP and $\sim$98.5 -- 99~{\AA}$^3$ according to WIEN2k.
%\item
\\ \indent $-$
The placement of the curve corresponding to the $\gamma$ polytype shows the largest
difference between the two calculations: it is the highest one according to VASP
but almost competing to the ground-state $\beta$ curve, according to WIEN2k.
In this relation it should be mentioned that the $\gamma$ phase was treated as
rhombohedral one in WIEN2k, according to its genuine primitive cell. However, in VASP
calculations we preferred to treat the $\gamma$ phase in the hexagonal setting,
(i.e., with triple amount of atoms in the unit cell), to make it more similar
to the other phases. Due to this compromise, the systematic error between the two codes
was presumably the largest in case of the $\gamma$ phase.
%\item
\\ \indent $-$
The $\varepsilon$ and the $\delta$ phases are consistently identified
as the highest one and the intermediate one, correspondingly, according to both calculations.
%\end{itemize} 

Taken together, this reveals a comforting level of agreement,
considering a very different background of calculation methods and the delicacy
of the differences to identify. 
%We tend to look at it as a justification of VASP results
%by comparison with an, a priori, superior method in what regards the precise
%total energy calculations for a given geometry, whereas the relaxed structure data
%might well come out more accurate in VASP, due to more sophisticated and less
%``human-dependent'' algorithm implemented in this code.

The second set of curves in the left panel of Fig.~\ref{fig:GaSe_EneVol},
marked ``$c$/$a$ relaxed'', which share the minimum points with their ``$c$/$a$ fixed''
counterparts but pass much more flat, represent the case where the structure was
fully relaxed (including also the internal coordinates) at each trial value of volume.
These curves could be used for extracting the bulk modulus (not discussed here
because of limited usefulness of this single parameter for the strongly anisotropic
systems under study). Corresponding calculations have not been done with WIEN2k,
because of too large amount of relaxations to be done effectively ``by hand''.

%\subsection{}
\begin{figure}[b]%
\includegraphics*[width=\linewidth]{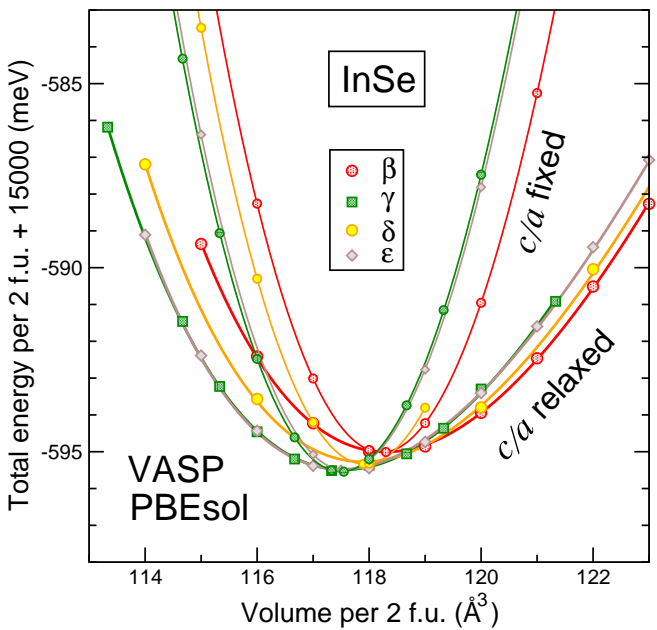}
\caption{\label{fig:InSe_EneVol}
Similar to the left panel of Fig.~\ref{fig:GaSe_EneVol}, for InSe polytypes.}
\end{figure}

Similar calculations (identification of the fully relaxed structure for each phase,
followed by ``rigid'' and ``$c$/$a$ relaxed'' volume scans) have also been done for
InSe; the results (from VASP calculations only) are shown in Fig.~\ref{fig:InSe_EneVol}.
Differently from the case of GaSe, one notes a small spread
of minimum positions -- from $\sim$117.5~{\AA}$^3$ (for $\gamma$ and $\varepsilon$,
which are almost degenerate and come about as ground states) to $\sim$118.5~{\AA}$^3$
for the $\beta$ phase which is now characterized by the highest-energy curve,
opposite to its atribution in GaSe. One notes however that at small negative pressure
(i.e., looking at the range of volumes larger than 120~{\AA}$^3$),
the $\beta$ phase would tend to dominate, and, in general, the whole system of 
$E(V)$ curves very faithfully recovers the order of phases established for GaSe.
A detailed discussion of this situation will be given elsewhere.
We note that the predominance of the $\gamma$ phase for InSe was long ago argued for
by Likforman and Guittard \cite{ComRenAcadSciC279-33}.

\section{I\lowercase{n}$_2$S\lowercase{e}$_3$ and 
G\lowercase{a}$_2$S\lowercase{e}$_3$ ordered-vacancies phases}
The (III)$_2$(VI)$_3$ stoichiometry restores the formal matching of cation / anion 
valences; the structure is compatible with tetrahedral environment 
if vacancies are allowed at 1/3 of the cation sites,
according to the formula (III)$_2\square$(VI)$_3$. Ye \emph{et al.} \cite{JJAP37-4264}
who discussed possible realisations of corresponding crystal structures
indicated different ways of ordering the vacancies in the 
underlying wurtzite structure, namely
the ``vacancies ordered in screw form'' (VOSF) and the
``layer structure phase'' (LSP). 
The VOSF unit cell is a $\sqrt{3}{\times}\sqrt{3}{\times}3$ multiplication of
the wurtzite one and contains $9{\times}4-6=30$ atoms, hence 6 formula units.
The corresponding space group can be either $P6_1$ (Nr. 169) or $P6_5$ (Nr. 170),
according to the screw being left-hand or right-hand.
The LSP phase possess the wurtzite lattice vectors in the basal plane
(without multiplication), and the vacancies are assembled into missing cation layers.
Instead of ``double layers'' (VI)-(III)-(III)-(VI), the system consists
of ``triple layer'' packages (VI)-(III)-(VI)-(III)-(VI), separated by 
van der Waals gaps between the opposing anion layers. Such structures are 
explained by Fig.~10 of Ye \emph{et al.} \cite{JJAP37-4264}.

\begin{table}[!t]
\caption{\label{tab:23-struc}
Lattice parameters (\AA) of Ga$_2$Se$_3$ and In$_2$Se$_3$ crystal structures
as optimized in calculations by VASP using the PBsol XC potential.
Experimental results \cite{JJAP37-4264} are indicated in brackets.
}
\medskip
\begin{tabular}{l*{4}{@{\hspace*{12pt}}r@{.}l}}
\hline 
\hspace*{-2mm}Polytype 
         & \multicolumn{2}{c}{\hspace*{-14mm}$a$} 
         & \multicolumn{2}{c}{\hspace*{-8mm}``wurtzite $a$''} 
         & \multicolumn{2}{c}{\hspace*{-14mm}$c$} 
         & \multicolumn{2}{c}{\hspace*{-8mm}``wurtzite $c/2$''} \\
\hline
\multicolumn{9}{c}{Ga$_2$Se$_3$} \\
VOSF & 6&73 & 3&88 & 18&90 & 3&15 \\
LSP1 & 4&04 & 4&04 & 28&89 & 3&21 \\
LSP2 & 4&04 & 4&04 & 28&96 & 3&22 \\
\multicolumn{9}{c}{In$_2$Se$_3$} \\
VOSF & 7&19 [7.14] & 4&15 & 20&23 [19.38] & 3&37 \\
LSP1 & 4&20 [4.00] & 4&20 & 29&39 [28.80] & 3&27 \\
LSP2 & 4&30        & 4&30 & 28&49         & 3&17 \\
\hline
\end{tabular}
\end{table}

Whereas such triple layers can be assembled in an infinite number of possible
sequences, as is generally the case with hexagonal polytypes, 
Ye \emph{et al.} \cite{JJAP37-4264}
distinguishes between the two simplest stackings, according to whether the
Se sheets within each triple layer follow the ``wurtzite-like'' sequence,
A-B-A, or ``zincblende-like'' one, A-B-C. 
Taken together, this yields two possibilities, shown in 
Fig.~10 of Ref.~\onlinecite{JJAP37-4264}:
\mbox{A-C-A$\cdots$B-A-B$\cdots$C-B-C$\cdots$} (``Model 1''), or
\mbox{A-B-C$\cdots$B-C-A$\cdots$C-A-B$\cdots$} (``Model 2'').
Indicating a place of a missing atom in parentheses, this would yield:
A-C-A$\cdot\cdot$(C)$\cdot\cdot$B-A-B$\cdot\cdot$(A)$\cdot\cdot$C-B-C$\cdot\cdot$(B)$\cdot\cdot$, or
A-B-C$\cdot\cdot$(A)$\cdot\cdot$B-C-A$\cdot\cdot$(B)$\cdot\cdot$C-A-B$\cdot\cdot$(C)$\cdot\cdot$.

Both these models for the layered phase have been tested in our calculation;
the optimized lattice parameters after VASP calculations are listed in
Table~\ref{tab:23-struc},
along with the measured data \cite{JJAP37-4264} for two structures of In$_2$Se$_3$.
For convenience of comparison between the phases,
also the lattice parameters ``reduced'' to the wurtzite setup are shown
($a/\sqrt{3}$ for VOSF, $c/6$ for VOSF and $c/9$ for LSP).
The relaxation procedure involved the internal coordinates; the discussion of
the corresponding trends in the relaxation around the vacancies is not included here.
As could be expected, the in-plane size of VOSF phase, that contains vacancies in every layer,
is more tight. Interestingly, the ``wurtzite $c$'' parameter of LSP phases
is (as intuitively expected) increased in Ga$_2$Se$_3$ but (a bit counter intuitively)
expanded in In$_2$Se$_3$, with respect to that of the VOSF phase. 
This is consistent with the fact that the LSP structures are hold together by
interaction between triple layers across the vdW gap, whereas the VOSF structures
are knitted by ``conventional'' covalent interactions in all directions.
The dispersive interactions,
\emph{a priori} not included in a conventional DFT calculation (also in those done with
PBEsol), seem to be stronger underestimated in Ga-Se systems than in In-Se ones.
Consequently, on insertion of ``vacancy layers'' the Ga$_2$Se$_3$ lattice
releases a bit, wereas the In$_2$Se$_3$ lattice becomes a bit tighter
due to a stronger covalent part in Se-Se interaction across the vdW gap.

For comparing relative stability of polytypes, it is not sufficient to discuss
the static total energies at equilibrium (so far available), but the study of
energy/pressure (or, energy/volume) curves would be due, in the spirit of as it 
is done above for the 1:1 phases. The corresponding analysis is beyond the scope
of this contribution.  

\begin{acknowledgements}
The PMMS (\emph{P\^ole Messin de Mod\'eli\-sa\-tion et de Simulation})
is gratefully acknowledged for providing us with computer time.
\end{acknowledgements}

%
%\bibliographystyle{pss}
%\bibliography{biblio}

\providecommand{\WileyBibTextsc}{}
\let\textsc\WileyBibTextsc
\providecommand{\othercit}{}
\providecommand{\jr}[1]{#1}
\providecommand{\etal}{~et~al.}

\end{document}